\documentclass[reprint,superscriptaddress,amsmath,amssymb,aps,prl]{revtex4-2}
\usepackage[utf8]{inputenc}
\usepackage{amsfonts}
\usepackage{graphicx}
\usepackage{xcolor}
\usepackage{graphicx}
\usepackage{dcolumn}
\usepackage{bm}
\usepackage{balance}
\usepackage{hyperref}
\usepackage[left=2cm,right=2cm,top=2cm,bottom=2cm]{geometry}
\hypersetup{
    colorlinks,
    linkcolor={red!50!black},
    citecolor={blue!50!black},
    urlcolor={blue!80!black}
}

\begin{document}

\title{Measurement of the dynamic charge susceptibility near the charge density wave transition in ErTe$_3$}

\author{Dipanjan Chaudhuri}
\email{dc36@illinois.edu}
\affiliation{Department of Physics, University of Illinois at Urbana-Champaign, Urbana, Illinois 61801, USA}
\affiliation{Materials Research Laboratory, University of Illinois at Urbana-Champaign, Urbana, Illinois 61801, USA}

\author{Qianni Jiang}
\affiliation{Department of Applied Physics, Stanford University, Stanford, CA 94305, USA}

\author{Xuefei Guo}
\affiliation{Department of Physics, University of Illinois at Urbana-Champaign, Urbana, Illinois 61801, USA}
\affiliation{Materials Research Laboratory, University of Illinois at Urbana-Champaign, Urbana, Illinois 61801, USA}

\author{Jin Chen}
\affiliation{Department of Physics, University of Illinois at Urbana-Champaign, Urbana, Illinois 61801, USA}
\affiliation{Materials Research Laboratory, University of Illinois at Urbana-Champaign, Urbana, Illinois 61801, USA}

\author{Caitlin S. Kengle}
\affiliation{Department of Physics, University of Illinois at Urbana-Champaign, Urbana, Illinois 61801, USA}
\affiliation{Materials Research Laboratory, University of Illinois at Urbana-Champaign, Urbana, Illinois 61801, USA}

\author{Farzaneh Hoveyda-Marashi}
\affiliation{Department of Physics, University of Illinois at Urbana-Champaign, Urbana, Illinois 61801, USA}
\affiliation{Materials Research Laboratory, University of Illinois at Urbana-Champaign, Urbana, Illinois 61801, USA}

\author{Camille Bernal-Choban}
\affiliation{Department of Physics, University of Illinois at Urbana-Champaign, Urbana, Illinois 61801, USA}
\affiliation{Materials Research Laboratory, University of Illinois at Urbana-Champaign, Urbana, Illinois 61801, USA}

\author{Niels de Vries}
\affiliation{Department of Physics, University of Illinois at Urbana-Champaign, Urbana, Illinois 61801, USA}
\affiliation{Materials Research Laboratory, University of Illinois at Urbana-Champaign, Urbana, Illinois 61801, USA}

\author{Tai C. Chiang}
\affiliation{Department of Physics, University of Illinois at Urbana-Champaign, Urbana, Illinois 61801, USA}
\affiliation{Materials Research Laboratory, University of Illinois at Urbana-Champaign, Urbana, Illinois 61801, USA}

\author{Eduardo Fradkin}
\affiliation{Department of Physics, University of Illinois at Urbana-Champaign, Urbana, Illinois 61801, USA}
\affiliation{Institute of Condensed Matter Theory, University of Illinois at Urbana–Champaign, Urbana, IL 61801, USA}

\author{Ian R. Fisher}
\affiliation{Department of Applied Physics, Stanford University, Stanford, CA 94305, USA}

\author{Peter Abbamonte}
\email{abbamont@illinois.edu}
\affiliation{Department of Physics, University of Illinois at Urbana-Champaign, Urbana, Illinois 61801, USA}
\affiliation{Materials Research Laboratory, University of Illinois at Urbana-Champaign, Urbana, Illinois 61801, USA}

\begin{abstract}

A charge density wave (CDW) is a phase of matter characterized by a periodic modulation of the valence electron density accompanied by a distortion of the lattice structure. The microscopic details of CDW formation are closely tied to the dynamic charge susceptibility, $\chi(q,\omega)$, which describes the behavior of electronic collective modes. Despite decades of extensive study, the behavior of $\chi(q,\omega)$ in the vicinity of a CDW transition has never been measured with high energy resolution ($\sim$meV). Here, we investigate the canonical CDW transition in ErTe$_3$ using momentum-resolved electron energy loss spectroscopy (M-EELS), a technique uniquely sensitive to valence band charge excitations. Unlike phonons in these materials, which undergo conventional softening due to the Kohn anomaly at the CDW wavevector, the electronic excitations display purely relaxational dynamics that are well described by a diffusive model. The diffusivity peaks around 250 K, just below the critical temperature. Additionally, we report, for the first time, a divergence in the real part of $\chi(q,\omega)$ in the static limit ($\omega \rightarrow 0$), a phenomenon predicted to characterize CDWs since the 1970s. These results highlight the importance of energy- and momentum-resolved measurements of electronic susceptibility and demonstrate the power of M-EELS as a versatile probe of charge dynamics in materials. 

\end{abstract}

\maketitle

Charge density waves (CDWs) are widespread in interacting electron materials and often coexist with other Fermi surface instabilities, such as magnetism or superconductivity \cite{gruner2018density}. Initially proposed for one-dimensional systems, CDW ground states have been observed across a range of strongly correlated materials with varying dimensionalities, including cuprate high-temperature superconductors \cite{hayden2023charge}, transition metal chalcogenides \cite{hwang2024charge}, transition metal bronzes, and organic conductors \cite{jerome2004organic}, among others \cite{gruner1988dynamics, chen2016charge}. A deep understanding of the behavior and underlying mechanisms of CDWs is crucial for advancing the study of quantum materials.

An expected thermodynamic signature of the onset of a CDW phase is a divergence in static charge susceptibility $\chi(q)$, which favors a periodic redistribution of charge in the ordered state \cite{gruner2018density}. 
This redistribution causes a gap to open in the single-particle density of states and is accompanied by a periodic distortion of the structural lattice driven by electron-phonon coupling. Although this phenomenon is most pronounced in one-dimensional systems where a Peierls instability due to Fermi surface nesting leads to logarithmic divergence \cite{gruner1988dynamics}, CDW phases have been widely observed in higher-dimensional systems. In these cases, nesting is imperfect and strongly dependent on the specific morphology of the Fermi surface \cite{johannes2006fermi,johannes2008fermi}.

The precise microscopic mechanism driving CDW mehavior in most materials remains elusive.  The nesting picture by itself is usually too simplistic, and effects such as temperature, scattering, and band velocity mismatch at the nesting wave vector play a crucial role in real materials \cite{johannes2008fermi, aebi2001search}. 
Correspondence between the CDW wave vector and the nesting properties of the Fermi surface are often difficult to discern \cite{rossnagel2011origin}. Other aspects, such as the momentum dependence of the electron-phonon interaction, have been invoked to quantitatively relate theory and experiments \cite{varma1983strong, johannes2006fermi, eiter2013alternative, flicker2015charge}. It is therefore crucial to experimentally measure the low-energy charge dynamics of a prototypical CDW material to understand the collective excitations that drive the CDW instability, and to determine whether, and how, a susceptibility divergence occurs.

Many of the essential properties of CDWs have been studied by a variety of experimental probes. The transport and thermodynamic anomalies associated with the CDW phase transition are widely observed in resistivity and specific heat measurements \cite{craven1977specific, gruner1988dynamics, wang1983charge, ru2006thermodynamic}, the periodic lattice distortion is measured with x-ray and neutron diffraction \cite{overhauser1971observability, hodeau1978charge, gruner1988dynamics, ru2008effect}, the CDW gap is probed through angle-resolved photo emission spectroscopy (ARPES) \cite{moore2010fermi, clerc2007fermi, rossnagel2011origin} and optical spectroscopy \cite{hu2011optical, degiorgi1991complete, perucchi2004optical}, and the soft phonon collective mode of a CDW has been observed with inelastic x-ray \cite{maschek2015wave, maschek2018competing, baron2020high} and neutron scattering experiments\cite{moncton1977neutron, sato1985neutron}. 

The behavior of the charge susceptibility itself, however, has never been measured. The reason is a shortage of experimental probes that measure the charge response at nonzero momentum, $q$, with meV energy resolution. Inelastic neutron scattering, for example, couples to matter through the nuclei or electron spins, so is only sensitive to phonons and spin excitations \cite{Balcar1989,Boothroyd2020}. Because x-rays couple to matter through the electron density, and most electrons in solids reside in core states, meV-resolved inelastic x-ray scattering (IXS), is also mainly sensitive to phonons, which modulate the electron density through displacement of the atomic cores \cite{vig2017measurement,ARCMP2024}. Resonant IXS (RIXS) is highly sensitive to valence band excitations, but instruments with $<10$ meV energy resolution are still a work in progress \cite{chaix2017dispersive, baron2020high}.

Here, we present measurements of the charge susceptibility of the prototypical CDW material ErTe$_3$ with momentum-resolved inelastic electron scattering (M-EELS). An advantage of scattering with electrons is that they couple to matter through the {\it charge} density, thereby directly measuring the dynamic charge susceptibility, $\chi''(q,\omega)$, for momenta ranging over multiple Brillouin zones \cite{ibach2013electron,vig2017measurement,ARCMP2024}. The current measurements were performed with an energy resolution of $\approx$ 5.6 meV (FWHM), which is sufficient to observe the low-energy charge excitations due to electrons near the Fermi surface involved in CDW formation. 
M-EELS has previously been successfully applied to valence band charge excitations in Bi$_2$Se$_3$ \cite{kogar2015surface}, TiSe$_2$ \cite{kogar2017signatures}, Bi$_2$Sr$_2$CaCu$_2$O$_{8+x}$ \cite{mitrano2018anomalous, husain2019crossover}, ZrSiS \cite{xue2021observation}, Sr$_2$RuO$_4$ \cite{husain2023pines}, and  SrTiO$_3$  \cite{kengle2023non}. M-EELS is therefore the ideal probe to finally investigate the charge susceptibility near a CDW transition. 

\begin{figure}%[tbhp]
\centering
\includegraphics[width=\linewidth]{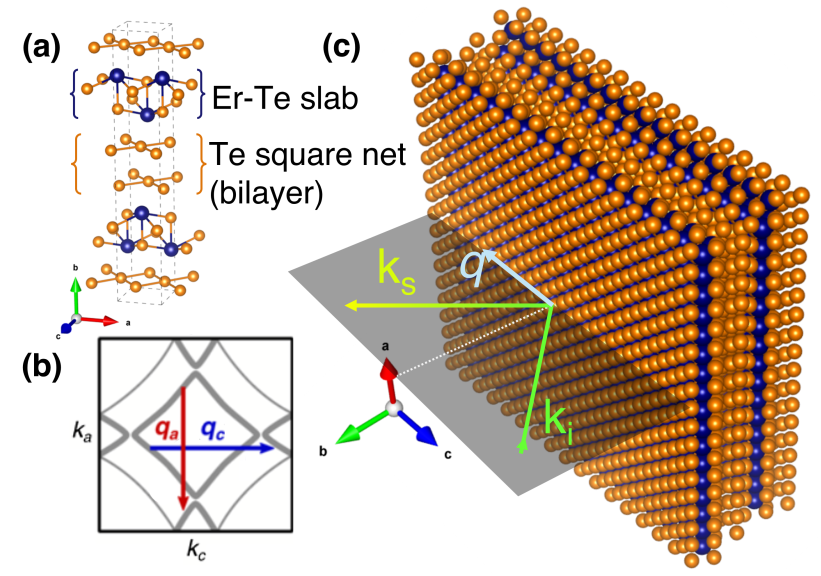}
\caption{(a) Crystal structure of ErTe$_3$ showing the corrugated Er-Te slabs and Te square nets. (b) Schematic of the Fermi surface showing the nesting vectors (adapted from Ref. \cite{straquadine2022evidence}). (c) Scattering geometry of the experiment with reference to the sample orientation. k$_i$ and k$_s$ represent momenta of the incident and scattered electron, respectively, and $q$ is the in-plane momentum transfer which is along $c$-axis.}
\label{fig:desc}
\end{figure}

The family of rare-earth tritellurides, $R$Te$_3$ ($R$ = La, Ce, Pr, Sm, Gd-Tm), are canonical quasi-2D CDW materials \cite{ru2006thermodynamic, ru2008effect, yumigeta2021advances}. They have a weakly orthorhombic crystal structure consisting of bi-layer, nominally square Te nets along the $ac$ planes that are separated by corrugated $R$Te slabs stacked along $b$-axis (see Fig.\ref{fig:desc}(a)). The precise CDW transition temperature depends on the chemical pressure exerted by the rare-earth atom. A second CDW, with lower transition temperature and a modulation orthogonal to the first, is observed when the rare-earth element is heavier than Dy \cite{ru2008effect}. Transport and x-ray diffraction studies have found that the CDWs in all these materials are incommensurate \cite{ru2008effect}. While phonon softening was observed in the vicinity of the CDW ordering wave vector \cite{maschek2015wave, maschek2018competing}, the electronic  excitations associated with the transition have not been studied. 

ErTe$_3$ is a representative of the series with two CDW transitions, at T$_{C1} \approx$ 267K and T$_{C2} \approx$ 159K with $q_{C1} \approx \left(5/7\right)c^*$ and $q_{C2} \approx \left(2/3\right)a^*$ respectively \cite{ru2008effect}. ARPES studies at low temperatures revealed a large gap at the Fermi surface, $\Delta_1 \approx 175$ meV along $c^*$, and a smaller gap, $\Delta_2 \approx 50$ meV was along $a^*$ \cite{moore2010fermi} (Fig.\ref{fig:desc}(b)). The magnitudes of these gaps are consistent with infrared spectroscopy \cite{hu2011optical}. While x-ray diffraction \cite{ru2008effect} and ARPES measurements \cite{moore2010fermi} were initially interpreted in terms of Fermi surface nesting alone \cite{yao2006theory}, detailed numerical simulations \cite{johannes2008fermi} along with Raman \cite{eiter2013alternative} and inelastic x-ray scattering \cite{maschek2015wave} studies have emphasized the importance of strongly momentum-dependent electron-phonon coupling in determining the CDW wavevector. The second, lower temperature CDW in the $R$Te$_3$ family is generally believed to be more subtle, since it does not show any signatures of phonon softening near the transition \cite{maschek2018competing}. We therefore focus our attention here on the higher transition in ErTe$_3$, which should serve as a canonical example of a CDW that exemplifies the behavior of the dynamic charge response at such a transition. 

\begin{figure}[hbt!]
\centering
\includegraphics[width=\linewidth]{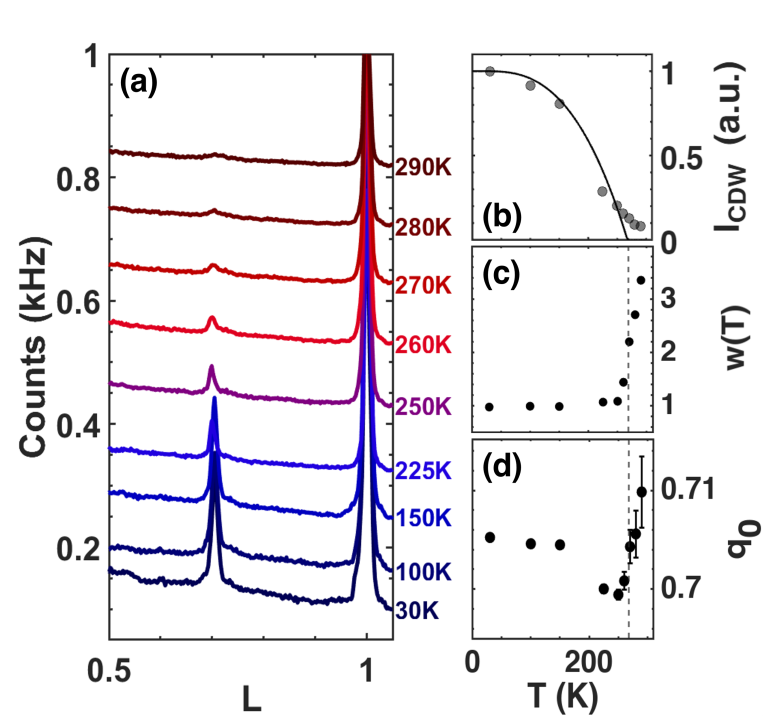}
\caption{\textbf{Elastic response:} (a) M-EELS spectra ($\omega \approx 0$) as a function of $q$ for different temperatures. (b) The integrated area under the CDW peak $q\approx q_0$ obtained from fits to \eqref{eq:static} along with the fit to mean-field interpolation formula, normalized to the value in the $T\rightarrow 0$. (c)  $w(t) = \gamma_{\text{CDW}}/\gamma_{(0,1)}$, the ratio of the width of the CDW satellite to that of the Bragg peak as a function of temperature. (d) Ordering wave vector $q_0$ as a function of temperature. Vertical gray dashed line in (c,d) denotes the T$_c$ obtained in fit in (b).}
\label{fig:static}
\end{figure}

Single crystal samples of ErTe$_3$, grown using self-flux method \cite{ru2006thermodynamic}, were cleaved perpendicular to the $b$-axis in ultra-high vacuum  prior to the M-EELS measurement. The cleaved samples were orientated such that the 50 eV incident electrons were scattered off along $bc$ plane with the in-plane momentum transfer, $q$, along the $c$ axis (see Fig.\ref{fig:desc}(c)). 

Elastic scattering from the CDW (energy loss $\hbar \omega \approx 0$), plotted as a function of $q$ for several temperatures in the vicinity of $T_{C1}$, is shown in \ref{fig:static}(a). Both the $(H,L)=(0,1)$ crystalline Bragg peak and the CDW satellite at $(H,L) \sim (0, 0.7)$ are visible in the scans. The width of the (0,1) Bragg peak is a measure of the overall momentum resolution of the experiment, which is better than 0.01 r.l.u. (see Supplemental Information).

The momentum lineshape of a CDW is known to be highly sensitive to 
disorder \cite{gruner1988dynamics}. 
Under perfect conditions, a CDW should have a power-law lineshape above its transition temperature, because of critical fluctuations near the transition, and become resolution-limited below the transition \cite{Holt2001,gruner2018density}. 
In the presence of disorder, however, a CDW can never exhibit true, long-ranged order in $d<4$ \cite{ImryMa1975}. Nie and coworkers \cite{nie2014quenched,lee2022generic} argued that, under such circumstances, the equal-time correlation function should instead be the sum of two terms, 
\begin{equation}
S(Q,T) = \sigma^2 G^2(Q,T) + T G(Q,T) \label{eq:lee}
\end{equation}
where $G(Q,T) = 1/[\kappa Q^2 + \mu(T)]$ is a Lorentzian, $\sigma$ represents the disorder strength, $Q=q-q_0$ is the momentum relative to the CDW wave vector, and the other parameters quantify the CDW stiffness and correlation length. The first term in \eqref{eq:lee} is truly static, while the second is dynamical and, in principle, could be eliminated by using an energy analyzer that perfectly isolated the elastic ($\omega = 0$) response. 

We found that the momentum lineshape of the CDW in ErTe$_3$ fit best to the single-Lorentzian expression, 
\begin{equation}
    I(q,T) = \dfrac{A(T)}{\left(q-q_{0}(T)\right)^2+\gamma_q^2(T)} + m(T) q + c(T)\label{eq:static}
\end{equation}
where $q_{0}(T)$ is the CDW ordering wavevector and the width, $\gamma_q(T)$, is proportional to the inverse correlation length (see Supplemental Information). The sum $m(T)q + c(T)$ represents a linear background whose meaning will be discussed further below. Surprisingly, adding an additional, Lorentzian-squared term did not improve the quality of the fits. This is surprising given that our measurement is energy-resolved, with an overall resolution of 5.6 meV, and should isolate the Lorentzian-squared term in \eqref{eq:lee}. We conclude that the CDW in ErTe$_3$ is influenced by disorder, but extremely weakly.

\begin{figure*}[hbt!]
\centering
\includegraphics[width=0.9\linewidth]{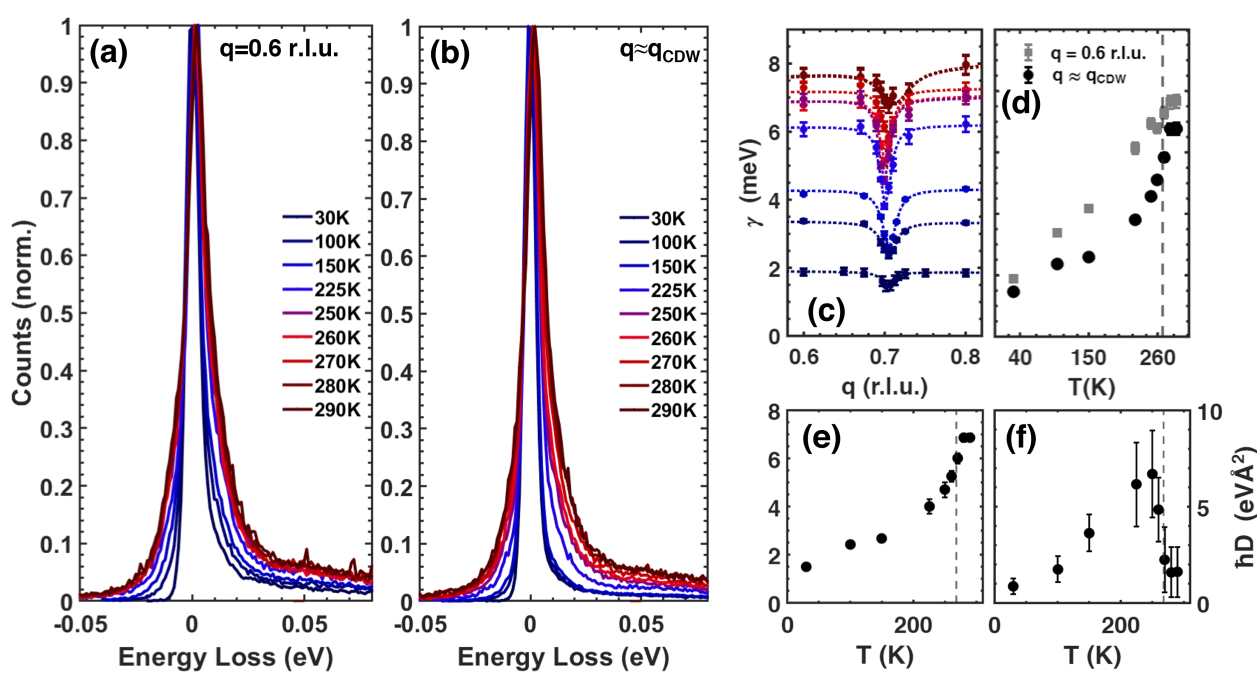}
\caption{\textbf{Dynamical response:} Low energy M-EELS spectra normalized to the maximum intensity for different temperatures for (a) $q=0.6 r.l.u.$ and (b) $q\sim q_0$. (c) $\gamma(q,T)$ extracted by fitting the loss spectra to \eqref{eq:loss_fit}. The dashed lines are a guide to the eye. (d) Temperature dependence of $\gamma(q,T)$ for $q = 0.6$ (r.l.u.) and $q\approx q_0$. (e) $\tau^{-1}$ and (f) $D$ obtained from fitting $\gamma(q,T)$ near $q_0$ to \eqref{diffusion}.}
\label{fig:dynamic}
\end{figure*}

The integrated intensity of elastic CDW reflection, $I_{\text{CDW}}(T)$, is plotted in Fig.\ref{fig:static}(b). $I_{\text{CDW}}(T)$ is proportional to the square of the order parameter, $\Delta(T)$, and a fit to the BCS interpolation formula $\Delta(T) = \Delta_0\tanh\left({1.74\sqrt{T_c/T-1}}\right)$, yields a $T_C = 268(7) K$ which is consistent with the previous x-ray diffraction measurements \cite{ru2008effect}. In the intermediate regime between 100-250K, the integrated intensity is smaller than the mean field prediction, which was also observed in the evolution of the single particle gap in ARPES \cite{moore2010fermi}. Vestiges of the CDW order is also observed above $T_c$, up to $T=$290 K, presumably due to pinning on disorder \cite{gruner1988dynamics}.

Fig.\ref{fig:static}(c) shows the ratio of the width of the CDW to that of the (0,1) Bragg peak, i.e., $w(t) = \gamma_{\text{CDW}}/\gamma_{(0,1)}$, as a function of temperature. Below, 225K the width of the CDW is nearly the same as that of the Bragg peak, indicating that it is limited by the instrumental resolution. 
This further supports the conclusion that disorder plays little role in the properties of this CDW. 
Above 225K, the CDW peak broadens, indicating that the CDW correlation length gets shorter as the temperature is increased through $T_c$. 
The CDW wavevector, $q_0(T)$, shows a non-monotonic dependence on temperature (Fig.\ref{fig:static}(d)). This behavior has also been observed in x-ray measurements of other tritellurides \cite{ru2008effect} and is a common characteristic of incommensurate CDWs. The exact value of $q_0$ often reflects a balance between factors such as the Fermi surface morphology, lattice pinning, the compressibility of uncondensed electrons, and coupling to hidden order parameters \cite{mcmillan1975,mcmillan1976,lee2022generic}. These behaviors of $q_0(T)$ suggests that the CDW in ErTe$_3$ is shaped by a combination of these effects.

Having established the structural properties of the CDW through elastic scattering, we now focus on its collective dynamics. Figures \ref{fig:dynamic}(a,b) present the temperature dependence of the M-EELS spectra at $q=0.6$ r.l.u. and near the CDW ordering wavevector, $q=q_0$. Across all temperatures and momenta, the spectra are dominated by a quasi-elastic response, with no evidence of dispersive collective excitations. This finding contrasts sharply with IXS studies, which reveal a softening of phonon modes near the CDW wavevector above $T_{C1}$ \cite{maschek2015wave, maschek2018competing}. Since M-EELS is more sensitive to valence band electronic excitations, while IXS primarily detects phonons, this suggests that the electronic excitations in ErTe$_3$ behave quite differently from its lattice excitations. This conclusion is further reinforced by the absence of phonon peaks in the low-energy optical conductivity of ErTe$_3$ \cite{hu2011optical}, an effect attributed to strong metallic screening.

Despite the absence of well-defined, propagating collective modes, the dynamical charge susceptibility shows a clear temperature dependence. A broadening of the quasi-elastic response is evident in the raw spectrum (Fig.\ref{fig:dynamic}(a,b)), suggesting that the CDW undergoes nontrivial, relaxational dynamics at an energy scale of  $\sim k_B T$. 

We parameterized this relaxational behavior using the Glauber model (also known as ``model A" \cite{chaikin1995principles}), which describes the diffusive dynamics of a non-conserved scalar order parameter.
In this model, the imaginary part of the susceptibility, $\chi''(q,\omega,T)$, is given by
\begin{equation}
    \chi''(q,\omega,T) = A(q,T)\dfrac{\omega}{\omega^2+\gamma^2(q,T)} \label{eq:modA}
\end{equation}
where $\omega$ represents the energy loss, $\gamma(q,T)$ is the relaxation rate that captures the diffusive dynamics, and $A(q,T)$ sets the overall scale. We then fit the M-EELS response using the expression
\begin{equation}
    I(q,\omega,T) =  V_{\text{eff}}^2(\omega,q) \; n(\omega,T) \;  \chi''(\omega,q,T) \label{eq:loss_fit}
\end{equation}
where $n(\omega,T)$ is the Bose occupation factor and $V_{\text{eff}}(\omega,q)$ is the Coulomb matrix element \cite{vig2017measurement}. This minimal model, with only two adjustable parameters ($A(q,T)$ and $\gamma(q,T)$) provides excellent fits to the M-EELS data 
(see Supplemental Information), confirming that the charge dynamics in ErTe$_3$ are predominantly relaxational.

The behavior of the relaxation rate in the vicinity of the CDW is summarized in Fig.\ref{fig:dynamic}(c), which shows $\gamma(q,T)$ as a function of $q$ for different temperatures. A pronounced minimum is observed near $q_0$, reflecting a prominent energy narrowing of the quasielastic line. This indicates critical slowing down of the CDW fluctuations near $q_0$ in the vicinity of the transition temperature. 
While this behavior is expected near a second-order phase transition, the scattering rate remains temperature-dependent across all momenta---even far from the CDW wavevector, $q_0$. This observation indicates that density fluctuations are present at all temperatures and length scales, beyond the conventional CDW fluctuations expected near $q_0$. 

However, the temperature dependence of the scattering rate differs significantly near the CDW wavevector compared to the rest of the Brillouin zone, as shown in Fig.\ref{fig:dynamic}(d). At $q=0.6$, the scattering rate decreases monotonically with temperature, resembling the behavior of the resistivity observed in transport experiments \cite{ru2008effect}. In contrast, near $q \sim q_0$, the scattering rate drops sharply as the sample is cooled through the CDW phase transition, then decreases more gradually as the temperature is further lowered deep into the ordered phase.

For momenta near the CDW ordering wavevector, the scattering rate has a roughly parabolic dependence on $q$. As a result, the CDW dynamics can be described by a diffusion model:
{
\begin{equation}
    \gamma(q,T) = \hbar\tau^{-1} + \hbar D(T) \;  (q-q_0)^2 \label{diffusion}
\end{equation}
}
where $\tau^{-1}$ represents pure dissipation and $D(T)$ is the diffusion constant. Figs. \ref{fig:dynamic}(e,f) show plots of $\tau^{-1}$ and $D(T)$, respectively. The dissipation rate, $\tau^{-1}$, varies quadratically with temperature, consistent with the behavior of a good Fermi liquid. Intriguingly, $D(T)$ peaks around $\sim$250 K, slightly below $T_c$ where a strong violation of the Wiedemann-Franz law was previously observed in thermal transport measurements \cite{Kountz2021}. This suggests that electronic CDW excitations propagate diffusively, with the diffusivity reaching its maximum just below $T_c$, when the CDW amplitude is large but still subject to significant thermal fluctuations. The characteristic diffusion length, $\lambda$, defined as $\lambda \sim \sqrt{D\;\tau}$, ranges between $\sim$ 2--10 lattice constants, comparable to that observed in the ``stripe-ordered'' phase La$_{2-x}$Ba$_x$CuO$_4$ $\left(x\sim\frac{1}{8}\right)$ with time-resolved x-ray scattering experiments \cite{Mitrano2019}. 

\begin{figure}[hbt!]
\centering
\includegraphics[width=0.8\linewidth]{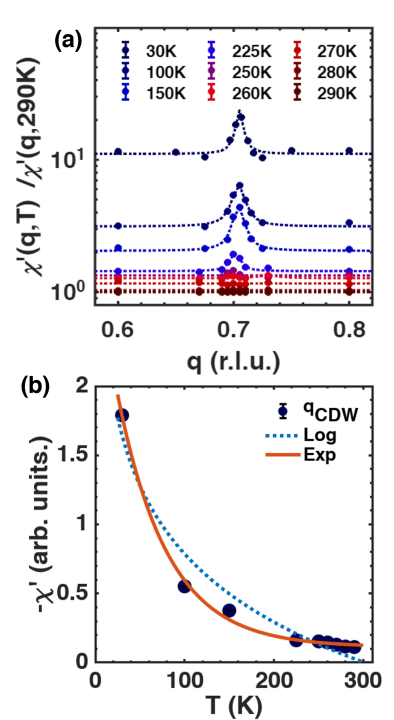}
\caption{(a) Real part of the charge susceptibility $\chi^{'}(q,\omega = 0,t)$ extracted from \ref{eq:chi'} normalized by the same for T=290K. The dashed lines are a guide to the eye. (b) Temperature dependence of $\chi^{'}(q_{0},\omega = 0,t)$, together with logarithmic and exponential fits.}
\label{fig:chi'}
\end{figure}

Finally, our analysis allows us to quantify the behavior of the static susceptibility, $\chi(q)$, which is traditionally expected to diverge near a CDW transition \cite{gruner2018density}. The static susceptibility is given by the real part of the complex susceptibility $\chi(q,\omega)$ evaluated at zero frequency. While this could be obtained via a Kramers-Kronig transformation of the data, we can also directly compute it from the real part associated with \eqref{eq:modA},
\begin{equation}
\chi'(q,\omega,T) |_{\omega=0} = A(q,T)/\gamma(q,T).  \label{eq:chi'}
\end{equation}
Fig.\ref{fig:chi'}(a) presents the static real part of the charge susceptibility as a function of $q$ at various temperatures, normalized to $\chi^{'}(q,0,290K)$. A peak is observed near $q_0$, which sharpens and increases in magnitude as the temperature decreases, suggesting a divergence associated with CDW formation. While such a peak is theoretically expected, this represents, to the best of our knowledge, the first experimental observation of this phenomenon.

One curiosity visible in Fig.\ref{fig:chi'}(a) is that the susceptibility does not diverge at $T_{C1}$, but continues to rise to the lowest temperature measured. This is a predicted property of the {\it unrenormalized} susceptibility \cite{gruner2018density}, not the fully dressed susceptibility measured in an experiment. One possible explanation for this behavior is that the transition influenced by disorder, which would cause the divergence to be rounded out. Another possibility is that only a small number of electrons contributing to $\chi(q,\omega)$ actually end up participating in the CDW order, in which case the bare and renormalized susceptibilities would be nearly the same. This is consistent with the ARPES measurement \cite{moore2010fermi} where the onset of the CDW order is only observed to gap a small fraction of the Fermi surface. The system overall remains metallic in the low temperature ordered phase \cite{ru2008effect}. Some insight may be gleaned from Fig.\ref{fig:chi'}(b), which shows the value of the susceptibility near $q_0$. Unlike the logarithmic divergence expected in 1D electron gas \cite{gruner2018density}, the curve shows exponential behavior, which could be a clue to the underlying mechanism. 

Finally, we note that, while the largest changes in the static susceptibility occur at $q_0$, the susceptibility shows significant temperature dependence at all momenta measured, which appear as a linear background in \eqref{eq:static}. This indicates that a substantial fraction of the valence electrons---even those not directly involved in CDW ordering---contribute significantly to the overall change in charge susceptibility. What underlying physics would drive such fluctuations everywhere in momentum space is an open question.   \balance

\begin{acknowledgments}
We thank Steven A. Kivelson and Anisha G. Singh for helpful discussions. M-EELS measurements were supported by the Center for Quantum Sensing and Quantum Materials, an Energy Frontier Research Center funded by the U.S. Department of Energy (DOE), Office of Science, Basic Energy Sciences (BES), under award DE-SC0021238. 
Q.J. and I.R.F (crystal growth and characterization) were supported by DOE BES contract DE-AC02-76SF00515. 
T.C.C (surface preparation) was supported by DOE BES contract DE-FG02-07ER46383. 
E.F. (theory) was supported by National Science Foundation grant DMR 2225920.
Q.J. was also partially supported by a Geballe Laboratory for Advanced Materials Postdoctoral Fellowship.
P.A. gratefully acknowledges support from the EPiQS program of the Gordon and Betty Moore Foundation, grant GBMF9452. 
\end{acknowledgments}

\bibliography{references}

\end{document}

\widetext
\begin{center}
\textbf{\large Supplemental Information: Measurement of the dynamic charge susceptibility near the charge density wave transition in ErTe$_3$}
\end{center}%%%%%%%%%% Merge with supplemental materials %%%%%%%%%%
%%%%%%%%%% Prefix a "S" to all equations, figures, tables and reset the counter %%%%%%%%%%
\setcounter{equation}{0}
\setcounter{figure}{0}
\setcounter{table}{0}
\setcounter{page}{1}
\makeatletter
\renewcommand{\theequation}{S\arabic{equation}}
\renewcommand{\thefigure}{S\arabic{figure}}
\renewcommand{\bibnumfmt}[1]{[S#1]}
\renewcommand{\citenumfont}[1]{S#1}
%%%%%%%%%% Prefix a "S" to all equations, figures, tables and reset the counter %%%%%%%%%%

The supplemental materials contains detailed fits to the static and the dynamical response discussed in the main text. The fits were obtained using the nonlinear least-squares solver in MATLAB. 

\section*{Elastic response}
\begin{figure} [hbt!]
\centering
\includegraphics[width=0.8\textwidth]{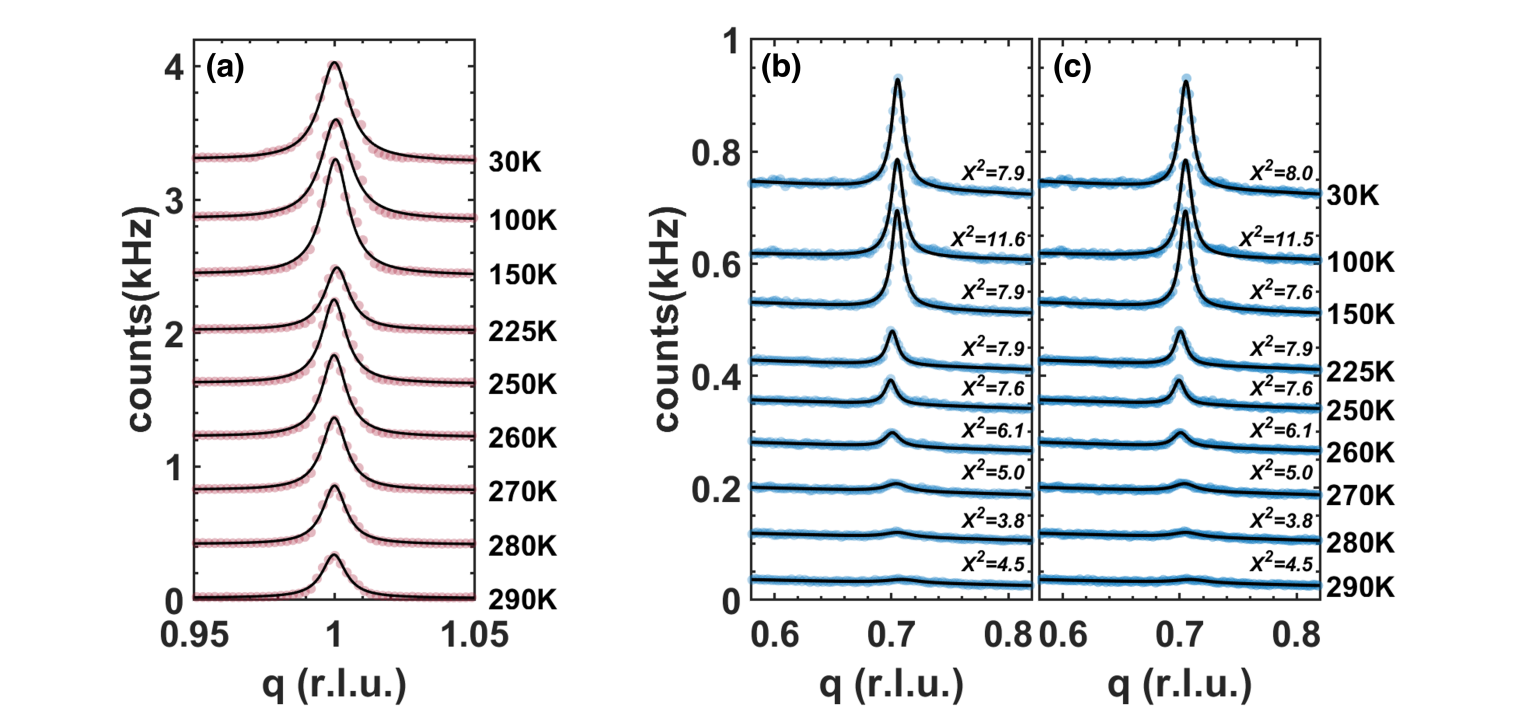}
\caption{Fits of the (a) Bragg peaks and (b) CDW satelites at each temperature to Eq. \eqref{eq:static_SI}. (c) Fits of the CDW satelites at each temperature to Eq. \eqref{eq:lee_SI}. The filled circles denote the data points and solid lines show the fitted curve. The spectra are plotted with vertical offset (0.4 for Bragg peaks, 0.075 for CDW) for clarity.}
\label{elastic_fit_SI}
\end{figure}

The Bragg peak and the CDW satellite at each temperature are fit to a Lorentzian form with a linear background given by, 
\begin{equation}
    I(q,T) = \dfrac{A(T)}{\left(q-q_{0}(T)\right)^2+\gamma_q^2(T)} + m(T) q + c(T)\label{eq:static_SI}
\end{equation}
where $q_{0}(T)$ is the center frequency ($q_{0}\sim0.7$ for CDW, $q_{0}\approx1$ for Bragg peak) and $\gamma_q(T)$ is the width. Lorentzian fits to the Bragg peak [Fig. \ref{elastic_fit_SI}(a)] are motivated by previous x-ray scattering experiments \cite{heiney1983x, stephens1984high}. As discussed in the main text, we found that the momentum-lineshape of the CDW satellites could be fit to a single-Lorentzian lineshape [Fig. \ref{elastic_fit_SI}(b)]. Furthermore, in Fig. \ref{elastic_fit_SI}(c), we show fits to the expression suggested by Nie and coworkers \cite{nie2014quenched,lee2022generic} for disordered systems (with a linear background), which is given by 
\begin{equation}
S(Q,T) = \sigma^2 G^2(Q,T) + T G(Q,T)  + m(T) q + c(T)\label{eq:lee_SI}
\end{equation}
where $G(Q,T) = 1/[\kappa Q^2 + \mu(T)]$ is a Lorentzian, $Q=q-q_0$ is the momentum relative to the CDW wave vector,  $\mu(T)$ is related to the CDW correlation length, $\sigma$ and $\kappa$ are temperature independent parameters that represent the disorder strength and CDW stiffness. We quantify the goodness of fit using Pearson's chi-squared test using the expression,
\begin{equation}
X^2 = \sum{\dfrac{\left(I_{exp}-I_{fit}\right)^2}{I_{fit}}}
\end{equation} 
where $I_{exp/fit}$ represent the experimental and fitted intensities respectively. Using this metric, we performed a global fit of the elastic response to Eq. \eqref{eq:lee_SI} at all temperatures and found the minima at $\sigma = 2.6$ and $\kappa = 2500$. Nevertheless, we observed that the fits were comparable to that obtained using Eq. \eqref{eq:static_SI} with lesser number of fit parameters. Hence, we conclude that the impact of disorder here is extremely weak.

\newpage

\section*{Dynamic response}
\begin{figure}[hbt!!]
\centering
\includegraphics[width=\textwidth]{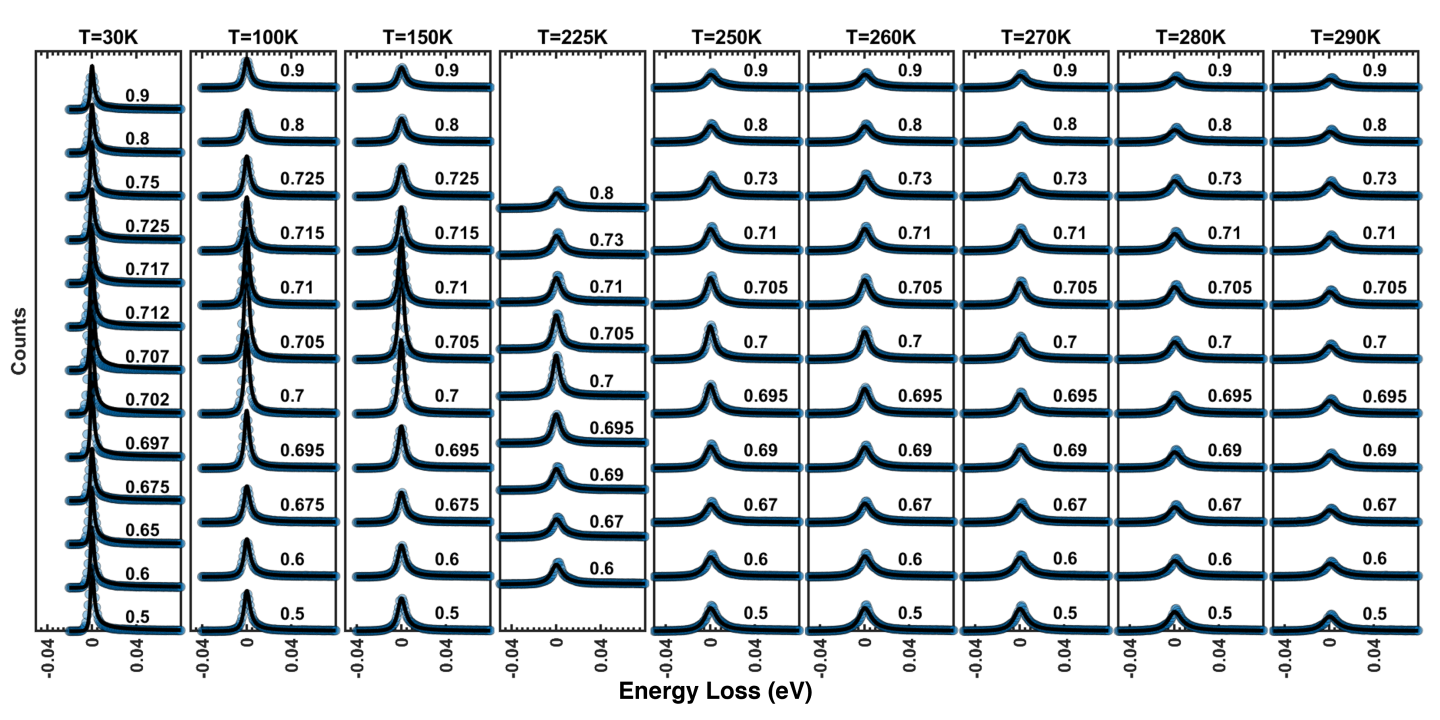}
\caption{Fits to the energy loss spectra at each temperature and momentum to Eq. \eqref{eq:loss_fit_SI}. The dots denote the data and solid lines show the fitted curve. The spectra are plotted with vertical offset for clarity.}
\label{inelastic_fit_SI}
\end{figure}

The M-EELS response can be modeled using the expression
\begin{equation}
    I(q,\omega,T) =  V_{\text{eff}}^2(\omega,q) n(\omega,T) \chi^{''}(\omega,q,T). \label{eq:loss_fit_SI}
\end{equation}
Here $n(\omega,T)$ is the Bose occupation factor 
\begin{equation}
    n(\omega,T) = -\dfrac{1}{\pi}\dfrac{1}{1-\exp{(-\hbar\omega/k_BT)}}
\end{equation}
and $V_{\text{eff}}(\omega,q)$ is the matrix element, given by the expression
\begin{equation}
    V_{\text{eff}}(\omega,q) = \dfrac{e^2/\epsilon_0}{q^2 + \left(k_i^z +k_s^z\right)^2}
\end{equation}
where $k_{i/s}$ refer to the incident and scattered electron momenta and can be calculated from the measurement geometry. For additional details, refer to \cite{vig2017measurement, chen2024consistency}. We use Glauber model for the dynamic charge susceptibility where the imaginary part, $\chi^{''}(q,\omega,T)$, is given by
\begin{equation}
    \chi^{''}(q,\omega,T) = A(q,T)\dfrac{\omega}{\omega^2+\gamma^2(q,T)}
\end{equation}
where $\gamma(q,T)$ is a scattering rate that captures the diffusive dynamics and $A(q,T)$ is an overall scale factor which is related to the real part of the charge susceptibility, $\chi^{'}(q,\omega,T)$. The fits are shown in fig. \ref{inelastic_fit_SI}.

\newpage

\section*{Scattering rate}
\begin{figure}[hbt!!]
\centering
\includegraphics[width=0.8\textwidth]{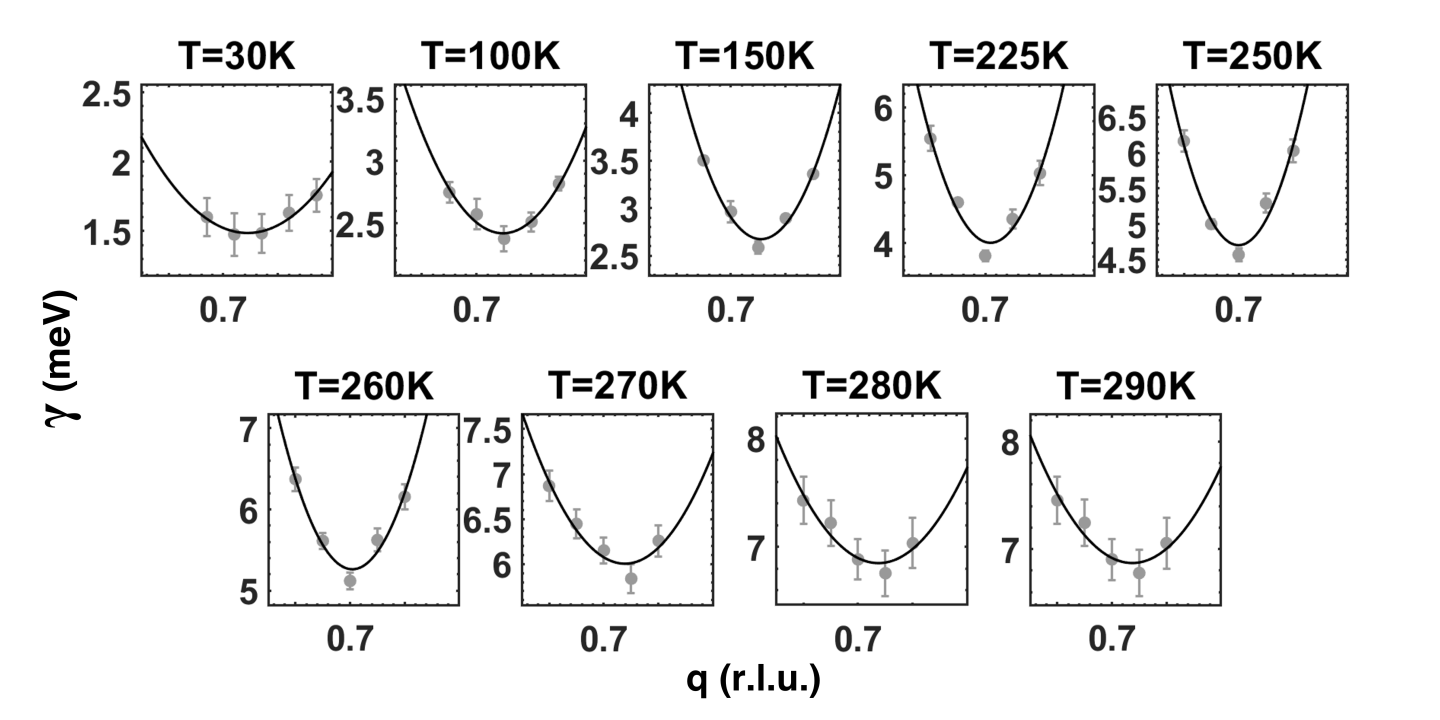}
\caption{Fits of the scattering rate extracted from the energy loss spectra to Eq. \eqref{eq:diffusion_SI} locally near the CDW wavevector. The dots denote the data and solid lines show the fitted curve.} \label{gamma_fit_SI}
\end{figure}

The momentum dependence of the scattering rate locally near the CDW ordering wavevector, can be approximated by a parabolic form based on simple diffusion model as
\begin{equation}
    \gamma(q,T) = \hbar\tau^{-1} + \hbar D(T) (q-q_0)^2 \label{eq:diffusion_SI}
\end{equation}
where $\tau^{-1}$ describes pure dissipation and $D(T)$ is the diffusion constant. The fits are showed in fig. \ref{gamma_fit_SI}.

\bibliography{references}